\documentclass[conference]{IEEEtran}
\usepackage{amsmath,graphicx}
\usepackage{psfrag} 

\usepackage[absolute]{textpos}
\newcommand{\copyrightstatement}{
    \begin{textblock}{15}(0.5,0.3)    
         \noindent
         \centering
         \textblockcolour{white}
         \footnotesize
         \copyright 2015 IEEE. Personal use of this material is permitted. Permission from IEEE must be obtained for all other uses, in any current or future media, including reprinting/republishing this material for advertising or promotional purposes, creating new collective works, for resale or redistribution to servers or lists, or reuse of any copyrighted component of this work in other works
    \end{textblock}
}

\begin{document}
%
\title{Estimating the HEVC Decoding Energy \\ Using the Decoder Processing Time}

\copyrightstatement

\author{\IEEEauthorblockN{Christian Herglotz, Elisabeth Walencik, and Andr\'e Kaup}
\IEEEauthorblockA{Multimedia Communications and Signal Processing\\
Friedrich-Alexander University Erlangen-N\"urnberg,
Cauerstr. 7, 91058 Erlangen, Germany\\ Email: \{ \ christian.herglotz, elisabeth.walencik, andre.kaup\ \}\ @FAU.de}}


%


\maketitle

\begin{abstract}
This paper presents a method to accurately estimate the required decoding energy for a given HEVC software decoding solution. We show that the decoder's processing time as returned by common C++ and UNIX functions is a highly suitable parameter to obtain valid estimations for the actual decoding energy. We verify this hypothesis by performing an exhaustive measurement series using different decoder setups and video bit streams. Our findings can be used by developers and researchers in the search for new energy saving video compression algorithms. 
\end{abstract}


\IEEEpeerreviewmaketitle

\section{Introduction}
During the last two decades, the demand for video services like streaming, recording, and coding increased rapidly. Nowadays, two thirds of the global internet traffic is related to video services \cite{cisco14}. Where in the early days research focused on reducing bit-rates while keeping the visual quality at a high level, today 
additional factors like the decoding energy obtain increased attention. This development can be explained by the fact that the typical use-case for video technologies switched from traditional desktop PCs to modern portable devices such as smartphones or tablet PCs. Due to the constrained dimension and resources of these devices, processing power and battery capacity has become a major limitation in the design of new algorithms. Hence, a low power consumption is a desirable property an algorithm should possess. 

A major obstacle in the search for energy efficient algorithms is the fact that power and energy measurements are complex and costly. In the usual case, a dedicated setup is needed that measures power and energy of the decoding system. At the same time it has to be ensured that no other process, like idle power or background system processes, disturb the measurements. Such setups have, e.g., been proposed in \cite{Li12}, \cite{Herglotz13}, \cite{Carroll10}, or \cite{Tseng10}. 


To overcome this drawback and obtain valid energy estimations without using such a setup, 
we show in this paper that the processing time as returned by common coding and development tools (C++ clock or UNIX time function) is highly linear correlated to the processing energy. This information can be exploited to optimize the energy efficiency of a given decoder solution without using a complicated and costly measurement setup as explained above. Hence, simple timing information of the decoding process suffices to prove the energetic superiority of a new algorithm implementation. 


In the following, we show experimentally that a proportional relation between time and energy exists for the case of software HEVC decoders. To prove this hypothesis, we tested different software solutions on different platforms, using single core as well as dual core processing, and different methods to determine the processing time. For each of these configurations we measured the decoding of a large variety of video bit streams. As a result we found that given the processing time $t_\mathrm{dec}$, the corresponding decoding energy $E_\mathrm{dec}$ can be estimated with a very high accuracy by 
\begin{equation}
E_\mathrm{dec} = P\cdot t_\mathrm{dec} + E_\mathrm{0}, 
\label{eq:energy_estimator}
\end{equation}
where $P$ is a linear factor and $E_\mathrm{0}$ a constant offset. 

This paper is organized as follows: Section \ref{sec:meas} explains the physical relation between time and energy and the measurement methods. Afterwards, Section \ref{sec:corr} introduces statistical basics that are used to determine the correlation between both values. Finally, Section \ref{sec:eval} lists the different decoding configurations that were evaluated, the set of video bit streams, and explains the results of the experiments. Section \ref{sec:concl} summarizes the conclusions.

\section{Time and Energy}
\label{sec:meas}
\begin{figure}
\centering
\psfrag{000}[r][c]{$0$}
\psfrag{001}[r][c]{$0.5$}
\psfrag{002}[r][c]{$1$}
\psfrag{003}[r][c]{$1.5$}
\psfrag{004}[r][c]{$2$}
\psfrag{005}[c][c]{}
\psfrag{006}[r][c]{$2.6$}
\psfrag{007}[r][c]{$3$}
\psfrag{008}[r][c]{$3.4$}
\psfrag{009}[r][c]{$3.8$}
\psfrag{010}[c][c]{}
\psfrag{011}[l][l]{Idle}
\psfrag{012}[l][l]{Decoding}
\psfrag{013}[b][c]{Power [W]}
\psfrag{014}[t][c]{Time [s]}
\includegraphics[width=0.45\textwidth]{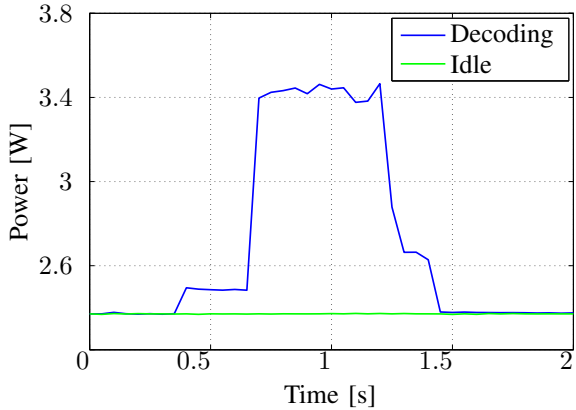}
\caption{Power consumption of the HM-decoder (system h1, see Section \ref{sec:eval}). The green line depicts the consumed power during idle mode (no active process) and the blue line the power during the decoding of a short video. The decoder starts at about $0.5$s and ends at $1.5$s. }
\label{fig:power}
\end{figure}

From a physical point of view, the consumed energy of a device is the time-integral over the power 
\begin{equation}
E = \int P(t)dt, 
\label{eq:eInt}
\end{equation}
where $E$ is the energy, $P(t)$ the instantaneous power, and $t$ the processing time. In general, the power $P(t)$ is a non-constant function of time. The power of an exemplary decoding process is depicted in Figure \ref{fig:power}. 

The blue line depicts the course of the power when the decoder is started at about $0.5$s and ends at $1.5$s. We can see that after an initialization phase and before the end a rather constant power of $3.4$W is required. Apparently, the power is not exactly constant. In spite of this observation we show experimentally that the approximation of a constant power is viable which simplifies (\ref{eq:eInt}) to 
\begin{equation}
E = \int P(t)dt \approx P\cdot t + E_\mathrm{0}, 
\label{eq:Eapprox}
\end{equation}
where $P=P(t)$ is the constant power. In addition, our measurements showed that the approximation is even more accurate when considering a potential offset energy $E_\mathrm{0}$. 
 Thus, we obtain two parameters that link the processing time (physical unit seconds [s]) with the decoding energy (physical unit Joule [J]): the mean power $P$ in Watts [W]  and the offset energy $E_\mathrm{0}$ in Joule [J]. 

To obtain valid and reliable results, we took great care to perform significant measurements. For the energy measurements, we constructed a dedicated measurement setup that is explained in Subsection \ref{sec:sec:energy_meas}. For the time measurements, we used C++-functions and functions that are included in typical Linux distributions as explained in Subsection \ref{sec:sec:time_meas}. 

\subsection{Energy Measurements}
\label{sec:sec:energy_meas}
\begin{figure}
\centering
\psfrag{U}[c][c]{$V_0$}
\psfrag{V}[c][c]{V}
\psfrag{D}[c][c]{DUT}
\psfrag{A}[c][c]{A}
\psfrag{P}[l][c]{Power meter}
\includegraphics[width=0.4\textwidth]{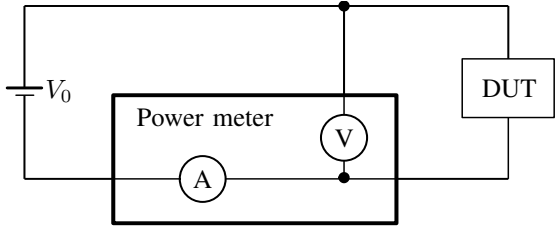}
\caption{Measurement setup with voltages source $V_\mathrm{0}$, decoding device (DUT), and power meter including an ampere meter (A) and a voltmeter (V). }
\label{fig:setup}
\end{figure}

To determine the energy consumed by a single process on a processing system, we used the test setup presented in Figure \ref{fig:setup}. 
The power meter is the LMG95 that internally calculates the consumed energy in a predefined time interval $T$. As device under test (DUT) we used different devices and device configurations as will be explained in Section \ref{sec:eval}. To separate the power consumption of the decoder from other consumers like LEDs or background processes we always performed two measurements: The first measurement was performed in idle mode to determine the energy consumption $E_\mathrm{idle}$ when no user process is active. During the second measurement the decoding process was executed so that the joint idle and decoding energy $E_\mathrm{all}$ is returned. As we are only interested in the pure decoding energy we subtract the first energy from the latter and obtain the decoding energy 
\begin{equation}
E_\mathrm{dec} = E_\mathrm{all} - E_\mathrm{idle}, 
\end{equation}
which visually corresponds to the area between the blue and the green curve in Figure \ref{fig:power}. 

Unfortunately, due to kernel processes and other factors such as ambient temperature and measurement inaccuracies, the measured energy 
was not constant for a fixed decoding system and video bit stream. Hence 
we performed multiple measurements until the mean measured energy met a specific statistical criterion. This criterion assures that the real mean value $\mu$ lies with a high probability inside a small region around the measured mean value $\hat{E} $. 

To test this criterion, we treat the energy as a random variable and perform a confidence interval test as proposed in \cite{Bendat71}. We perform $M$ measurements for each bit stream and each configuration, and calculate the mean energy
\begin{equation}
 \hat{E}_{\mathrm{dec},n} = \frac{1}{M}\sum_{m=1}^{M} E_{\mathrm{dec,}n,m}, 
 \label{eq_mean}
\end{equation}
where $E_{\mathrm{dec,}n,m}$ is the measured energy of the $m$-th measurement and $n$ the bit stream index. Afterwards, we calculate the size of the confidence interval around the measured mean value in which the real mean value is located within probability $\alpha$
\begin{equation}
  \hat{E}_{\mathrm{dec,}n} - \frac{\sigma}{\sqrt{M}}\cdot t_\alpha (M-1) < \mu  < \hat{E}_{\mathrm{dec,}n} + \frac{\sigma}{\sqrt{M}}\cdot t_\alpha (M-1),  
  \label{eq:confInt}
\end{equation}
where $\sigma$ denotes the standard deviation of the measurement and $t_\alpha$ the critical t-value of the student t distribution. After each measurement we test whether the current interval boundaries are below a constant fraction of the current mean value 
\begin{equation}
  \frac{\sigma}{\sqrt{M}}\cdot t_\alpha (M-1) < 0.01 \cdot \hat{E}_{\mathrm{dec,}n}
  \label{eq:boundary1}
\end{equation}
and repeat the measurement until this criterion is met. Thus we make sure that the real mean value of the current measurement lies within a probability of $\alpha=99\%$ inside the small interval around the measured mean value. 

\subsection{Time Measurements}
\label{sec:sec:time_meas}
We tested different timing functions that are easy to use and that allow measurements on any processing system, independent from the hardware and the operating system. The functions are the following: 

\begin{itemize}
\item C++ \texttt{clock()}-function: This function is part of the standard C++-library \texttt{<time.h>} \cite{Deitel98} and commonly used to determine the processing time of algorithm implementations. The function returns the current date in seconds. By subtracting the value at the beginning of the process from the value at the end we obtain the time elapsed during the process. 
\end{itemize}
The UNIX specific command \texttt{time} allows for measuring different time values and can only be used on UNIX-based systems \cite{stevens99}. 
The tested values are the following: 
\begin{itemize}
\item real time: This value is the time that elapsed from the start until the end of the process. It is comparable to the value returned by the C++ \texttt{clock()} function. 
\item user time: This value is the amount of CPU time attributed to the process. Hence, simultaneously running background processes do not influence this time. On multi-core systems, this function adds up the times of all CPUs that are involved in the process. 
\end{itemize}
Note that the C++ \texttt{clock()} time and the UNIX real time may vary because background processes can interfere with this measurement. Hence, we also performed the statistical test explained in Section \ref{sec:corr} for these values. 

\section{Statistical Correlation}
\label{sec:corr} 
As a result of the measurements we obtain energy-time pairs for each decoding configuration. To determine the relation between the values for each configuration, we performed a linear correlation analysis as proposed in \cite{Bendat71}. The correlation coefficient for time and energy is calculated as
\begin{equation}
  r_\mathrm{tE} = \frac{\sum_{n=1}^{N}(\hat{E}_{\mathrm{dec,}n}- \hat{E}_\mathrm{dec})\cdot(\hat{t}_{\mathrm{dec,}n}- \hat{t}_\mathrm{dec})}{\left[  \sum_{n=1}^{N}(\hat{E}_{\mathrm{dec,}n}- \hat{E}_\mathrm{dec})^2 \cdot     \sum_{n=1}^{N}  (\hat{t}_{\mathrm{dec,}n}- \hat{t}_\mathrm{dec})^2         \right]^\frac{1}{2}}, 
  \label{eq:linCorr}
\end{equation} 
where $n$ is the bit stream index and $N$ the total number of bit streams. $\hat{E}_\mathrm{dec}$ and $\hat{t}_\mathrm{dec}$ denote the mean values over all $N$ bit streams. 

As this analysis returned values close to one for each configuration, which corresponds to a very high linear correlation, we decided to calculate the linear regression of the samples to estimate the energy consumption $E_\mathrm{dec}$ in dependence of the processing time $t_\mathrm{dec}$ as shown in (\ref{eq:energy_estimator}). The slope of this line (the so-called regression coefficient) is calculated by 
\begin{equation}
	P = \frac{\sum_{n=1}^{N}(\hat{E}_{\mathrm{dec,}n}-\hat{E}_\mathrm{dec})\cdot(\hat t_{\mathrm{dec,}n}-\hat{t}_\mathrm{dec})}{ \sum_{n=1}^{N}  (\hat t_{\mathrm{dec,}n}-\hat{t}_\mathrm{dec})^2 }. 
	\label{eq:slope_regr}
\end{equation}
This coefficient can be interpreted as the mean power that is required for the decoding process. 

Afterwards, the offset $E_\mathrm{0}$ can be calculated by  solving (\ref{eq:energy_estimator}) for $E_\mathrm{0}$, where energy and time values are replaced by their arithmetic means:
\begin{equation}
  E_\mathrm{0} = \hat{E}_\mathrm{dec} - P \cdot \hat{t}_\mathrm{dec}. 
  \label{eq:offset_regr}
\end{equation}
Such a model was created for every decoder configuration as evaluated in the next section. 

\section{Evaluation}
\label{sec:eval}
In this section we show that the proposed linear regression is applicable for energy estimation in the field of HEVC software decoding. Hence, the different decoder configurations are presented in the first subsection, followed by the set of video bit streams. The results subsection shows the validity of the proposed linear model. 

\subsection{Decoder Configurations}
To evaluate the model we tested two hardware platforms: A Pandaboard \cite{Panda} and an FPGA featuring a softcore processor. 
The Pandaboard has a smartphone-like architecture featuring an OMAP4430 SoC with an ARMv7 dual core processor. As an operating system we used Ubuntu 12.04 and switched off LEDs and unnecessary system processes. The board was controlled via RS232. The decoder configurations that were measured for this device (l1 to h3) are listed in Table \ref{tab:configs}. The output of the decoding processes (the raw bit stream) was discarded by sending it to the null device (/dev/null) so that no stall occurs.

\begin{table}[t]
\renewcommand{\arraystretch}{1.3}
\caption{Decoder configurations for the Pandaboard (ID l1 to h3) and the SPARC processor on the FPGA (ID h4). Column RL denotes the runlevel of the operating system and column Optim. denotes the gcc compiler optimization flag that was used to compile the decoder. }
\label{tab:configs}
\begin{center}
\footnotesize{\begin{tabular}{c|l|c|c|c|c}
\hline
ID & Software & RL & Optim.  & Cores & Timing method \\
\hline
l1 & libde265 \cite{libde} & 1 & -o3 & Single & C++ \texttt{clock()} \\
f1 & FFmpeg \cite{FFmpeg} & 1 & - & Single & UNIX real time \\
f2 & FFmpeg  & 1 & - & Dual & UNIX real time \\
f3 & FFmpeg & 1 & - & Dual & UNIX user time \\
h1 & HM-13.0 \cite{HM-13.0} & 1 & -o3 & Single & C++ \texttt{clock()} \\
h2 & HM-13.0  & 2 & -o3 & Single & C++ \texttt{clock()} \\
h3 & HM-13.0  & 1 & -o0 & Single & C++ \texttt{clock()} \\
\hline
h4 & HM-11.0 & - & -o3 & Single & C++ \texttt{clock()} \\
 \hline
\end{tabular}}
\end{center}
\end{table}

The HM-software as well as libde265 were compiled on the Pandaboard while we used a static build for FFmpeg. Different runlevels and compiler optimization flags were tested to see if background processes and different processor loads, respectively, influence the measurements.

In a further measurement we tested whether the linear model is valid for a completely different hardware setup (ID h4). Therefore, we cross compiled the HM-11.0 software for a SPARC processor and executed the decoder on a Leon3 soft-IP core on an Altera Terasic DE2-115 FPGA board. The execution was performed bare-metal, i.e. without an underlying operating system. The board was controlled using GRMON debugging tools \cite{grmon2} and the time was measured using the C++ \texttt{clock()} function. 

\subsection{Video Bit Streams}
To show that the model is independent from the type of video bit stream we measured the decoding of $120$ different test video bit streams for each configuration. They were encoded using the HM-13.0 reference software using the four encoder configurations intra, lowdelay, lowdelay\_P, and randomaccess. QP was set to $10$, $32$, and $45$. The ten source videos listed in Table \ref{tab:eval_vids} were taken from the HEVC test set. The frame number was chosen arbitrarily. 

\begin{table}[t]
\renewcommand{\arraystretch}{1.3}
\caption{Properties of evaluated videos. Videos were taken from the HEVC test set and were encoded with HM-13.0 using standard encoder configurations intra, lowdelay\_P, lowdelay, and randomaccess. QP was set to 10, 32, and 45.  }
\label{tab:eval_vids}
\vspace{-0.3cm}
\begin{center}
\footnotesize{\begin{tabular}{l|c|c|r}
\hline
Name & Class & Resolution & No. frames \\
\hline
PeopleOnStreet & A & $2560\times1600$ & 16 \\
Traffic & A & $2560\times1600$ & 16 \\
Kimono & B & $1920\times1080$ & 16 \\
RaceHorses & C & $832\times480$ & 16 \\
BasketballPass & D & $416\times240$ & 40 \\
BlowingBubbles & D & $416\times240$ & 40 \\
BQSquare & D & $416\times240$ & 40 \\
RaceHorses & D & $416\times240$ & 40 \\
vidyo3 & E & $1280\times720$ & 16 \\
SlideEditing & F & $1280\times720$ & 30 \\
 \hline
\end{tabular}}
\end{center}
\end{table}
Unfortunately, due to the restricted size of the RAM on the FPGA, we had to check a different set of videos for configuration h4. For this case, we only tested $83$ different bit streams from class B to F with a lower number of frames but the same QPs and encoder configurations.

\subsection{Results}
\begin{figure}
\centering
\psfrag{000}[r][c]{$0$}
\psfrag{001}[r][c]{$20$}
\psfrag{002}[r][c]{$40$}
\psfrag{003}[r][c]{$60$}
\psfrag{004}[r][c]{$80$}
\psfrag{005}[r][c]{$100$}
\psfrag{006}[r][c]{}
\psfrag{007}[r][c]{$10$}
\psfrag{008}[r][c]{$20$}
\psfrag{009}[r][c]{$30$}
\psfrag{010}[r][c]{$40$}
\psfrag{011}[r][c]{$50$}
\psfrag{012}[r][c]{$60$}
\psfrag{013}[l][l]{Regression}
\psfrag{014}[l][l]{Measurement}
\psfrag{015}[c][t]{Decoding Energy [J]}
\psfrag{016}[c][b]{Processing Time [s]}
\includegraphics[width=0.49\textwidth]{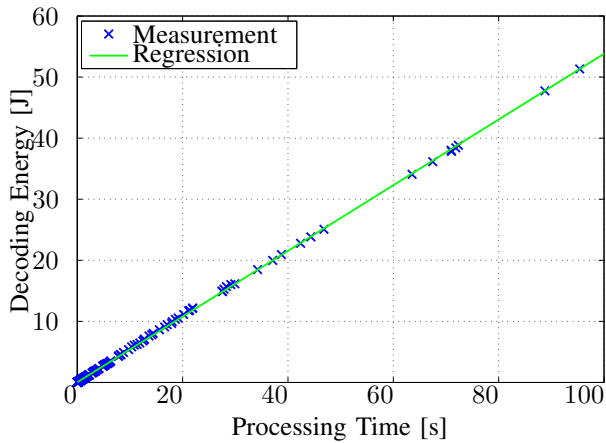}
\caption{Distribution of time and energy for different video bit streams decoded with HM-13.0 on the Pandaboard (configuration h1). Each marker depicts one bit stream and the straight line the result of the linear regression. }
\label{fig:regression}
\end{figure}

To show the behavior of the decoding energy in relation to the processing time, Figure \ref{fig:regression} shows all time-energy pairs for decoder configuration h1. 

We can see that all markers, where each marker depicts the decoding of one of the video bit streams, approximately lie on a straight line. To further prove this observation, we also plot the result of the linear regression (green line). As the other configurations show similar results, we summarize the correlation coefficient and the regression coefficients in Table \ref{tab:corrCoeffs}. 

\begin{table}[t]
\renewcommand{\arraystretch}{1.3}
\caption{Correlation coefficients and linear regression coefficients for all tested decoder configurations. }
\label{tab:corrCoeffs}
\vspace{-0.3cm}
\begin{center}
\footnotesize{\begin{tabular}{l||r||r|r}
\hline
Configuration & Correlation $r_{tE}$ & Slope $P$ [W] & Offset $E_\mathrm{0}$ [J] \\
\hline
l1 & $0.999946$ & $0.5130$ & $0.0534$ \\
f1 & $0.999931$ & $0.5236$ & $0.0342$ \\
f2 & $0.998315$ & $0.8964$ & $-0.1240$ \\
f3 & $0.999571$ & $0.5293$ & $0.0546$ \\
h1 & $0.999937$ & $0.5377$ & $0.0480$ \\
h2 & $0.999917$ & $1.0456$ & $-0.2186$ \\
h3 & $0.999922$ & $0.4732$ & $0.1040$ \\
h4 & $0.999660$ & $0.3351$ & $0.2619$ \\
 \hline
\end{tabular}}
\end{center}
\end{table}

We can see that all correlations are very close to one. Main differences can be found in the mean power $P$. 
An interesting observation is, for instance, that all processes on the Pandaboard in runlevel 1 (configurations h1, l1, f1) require approximately the same power. Only the o0-optimized configuration h3 requires less power which can be explained by a lower processor capacity utilization. In contrast, using dual core processing (configuration f2) the power rises significantly due to the second core that is active. 

As a third observation, the mean power using user time (configuration f3) in dual core processing is close to the power in runlevel 1 for single core processing which can be explained by the fact that both CPU times are added and the mean yields the average CPU processing power. 

Furthermore, it is striking that the power in runlevel 2 (configuration h2) is much higher than in runlevel 1 which is caused by the additional background processes. 

Finally, we can see that the offsets $E_\mathrm{0}$ are unequal zero. To explain this effect we performed another test series for very short video bit streams (one or two frames of class D resolution). We could see that for these bit streams non-linear effects occur, where the first energy-time pairs appear close to the origin and then asymptotically approach the regression curve. 

\section{Conclusion}
\label{sec:concl}
To sum up, we have shown that for the case of software video decoding, it is a valid approach to use the processing time as an estimator for the processing energy. As the relation is linear, it is even not necessary to know the values of the coefficients linking energy and time. Hence, the minimization of the processing time is equivalent to the minimization of the processing energy, which is a very beneficial information for software developers and video coding researchers.


\section*{Acknowledgment}
This work was financed by the Research Training Group 1773 ``Heterogeneous Image Systems'', funded by the German Research Foundation (DFG).

\bibliographystyle{IEEEtran}
\bibliography{IEEEabrv,literatureNeu}

\end{document}